\documentclass[journal]{IEEEtran}

\usepackage{amsmath,amssymb}
\usepackage{algorithm}
\usepackage{algorithmic}
\usepackage{booktabs}
\usepackage{graphicx}
\usepackage{microtype}
\usepackage{multirow}
\usepackage{tabularx}
\usepackage{url}
\usepackage{cite}
\usepackage[hidelinks]{hyperref}

\newcommand{\agent}{\textsc{Agent}}
\newcommand{\fabric}{\textit{Agent-Native Metamorphic Communication Fabric}}

\DeclareMathOperator{\diag}{diag}

\begin{document}

\title{Agent-Native Metamorphic Communication Fabric}

\author{Jienan~Chen,~\IEEEmembership{Senior~Member,~IEEE}
\thanks{Jienan Chen is with the National Key Laboratory of Science and Technology on Communications, University of Electronic Science and Technology of China (UESTC), No. 2006 Xiyuan Avenue, West Hi-Tech Zone, Chengdu 611731, Sichuan, P. R. China (e-mail: jesson.chen@outlook.com).}}

\maketitle

\begin{abstract}
Communication intelligence is undergoing two related transitions: algorithm development is moving from predominantly manual model-based design toward foundation-model-assisted generation and evaluation, while deployed systems are moving from offline optimization toward agent-driven online decision and guarded deployment. Existing data-driven or large-language-model-assisted methods remain primarily design-time tools and cannot by themselves cover every future combination of service intent, channel, spectrum, and hardware state. This paper proposes the \fabric, a closed-loop architecture in which an agent observes the operating state, selects or generates an explicit communication candidate, executes it in a digital twin, applies hard feasibility gates, and deploys it with monitoring and fallback. Three levels define the allowed scale of change: Level~1 adjusts parameters while preserving the algorithm; Level~2 switches and configures receiver algorithms while preserving the protocol and waveform; and Level~3 reconfigures the waveform or waveform--multiple-access chain while preserving the service contract and safety interface. Simulations demonstrate all three levels. Level~1 improves rate, channel tracking, or quantization energy under fixed topologies. Level~2 selects three-iteration weighted Jacobi, five-iteration diagonally preconditioned conjugate gradient, and direct MMSE in favorable, intermediate, and harsh MIMO regimes, respectively; the analytical hardware proxy predicts up to 67.1\% energy and 73.3\% latency reduction relative to direct MMSE. Level~3 selects CP-OFDM, SC-FDMA, OTFS, filtered OFDM, and SCMA-over-OFDM across five representative operating regimes and forms continuous switching boundaries under Doppler, spectrum-contiguity, load, and RF-power sweeps. The results establish a minimum viable mechanism for verifiable runtime communication adaptation rather than unconstrained end-to-end learning or arbitrary online code mutation.
\end{abstract}


\section{Introduction}
\label{sec:introduction}

\IEEEPARstart{W}{ireless} systems have traditionally been engineered around a small number of stable assumptions: a standardized frame structure, a finite set of waveforms, and adaptation rules designed and validated before deployment. This process has enabled robust and interoperable communication at scale. It is increasingly strained, however, by autonomous and heterogeneous services whose intent, mobility, reliability target, spectrum opportunity, and hardware budget can change on operational timescales. The IMT-2030 framework and broader sixth-generation visions explicitly couple communication with distributed intelligence, sensing, and heterogeneous computation \cite{iturM2160,rappaport2019wireless,saad2020vision}. A remaining question is whether the communication system itself can change form when its operating context changes.

The evolution of communication design can be summarized through three paradigms. In the model-driven paradigm, engineers construct a channel and system model, derive an algorithm, map it to software and hardware, and repeatedly validate the implementation. In the data-driven paradigm, a neural transmitter, receiver, or intermediate module is trained from samples, but the learned structure is still normally fixed after offline training \cite{oshea2017deep,aoudia2018model,cammerer2023neural}. More recently, foundation models have become algorithm-development agents: they can connect literature, mathematical models, code, execution feedback, and evaluators to iteratively improve an implementation. AlphaEvolve demonstrates this pattern for general scientific and algorithmic discovery \cite{novikov2025alphaevolve}, and the AI Telco Engineer applies related evolutionary coding agents to wireless receiver design \cite{aoudia2026aite}. These systems materially expand the scale of offline algorithm search, yet a candidate produced offline cannot anticipate every future service and channel combination.

\begin{figure*}[t]
    \centering
    \includegraphics[width=0.97\textwidth]{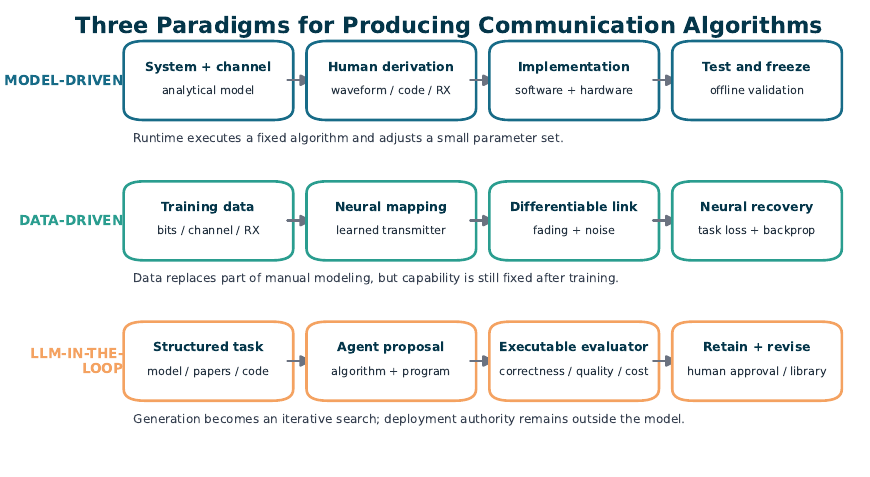}
    \caption{Evolution of communication-algorithm production. Model-driven and data-driven workflows normally freeze the principal algorithmic artifact before deployment. LLM-in-the-loop workflows make proposal and executable evaluation iterative, but do not by themselves provide runtime deployment authority.}
    \label{fig:design-paradigms}
\end{figure*}

The transition pursued in this paper is therefore not simply a stronger design assistant. It is a runtime communication agent that can observe state, modify an approved part of the transceiver, evaluate the modification with executable evidence, and deploy it only when explicit safety gates pass. The distinction is important. An agent-native system need not replace the transceiver by an opaque neural model, nor must a large language model sit in the real-time data path. The agent can combine a language-model proposer, a deterministic optimizer, a registered algorithm library, and a digital-twin evaluator. What makes the system agent-native is that observation, candidate formation, verification, decision, deployment, and fallback are part of the operating loop rather than an external design workflow.

The motivating question is deliberately simple: must a short-range link always use OFDM? A universal answer is not meaningful. CP-OFDM can be preferable for flexible scheduling or continuous wideband spectrum, SC-FDMA can be preferable when power-amplifier efficiency dominates, OTFS can become useful under strong Doppler, and a receiver may safely replace direct matrix factorization with a finite iterative solver when the channel matrix is well conditioned. The desired communication form therefore depends jointly on task, channel, spectrum, load, and hardware constraints.

This paper proposes the \fabric{} as an architectural and experimental framework for this transition. The word \emph{metamorphic} denotes controlled change of communication form under explicit invariants; it does not denote unconstrained mutation. The framework defines three levels of online autonomy:
\begin{enumerate}
    \item \textbf{Level 1 -- parameter adaptation:} the protocol, waveform, and algorithm topology remain fixed, while the \agent{} selects parameters such as MCS, pilot density, equalizer regularization, or quantization precision;
    \item \textbf{Level 2 -- receiver algorithm adaptation:} the protocol, waveform, and receiver objective remain fixed, while the \agent{} selects and configures a numerical solution path such as weighted Jacobi, diagonally preconditioned conjugate gradient (PCG), or direct MMSE; and
    \item \textbf{Level 3 -- waveform and multiple-access adaptation:} the service contract, regulatory constraints, and safety interface remain fixed, while the \agent{} selects a signal representation or waveform--multiple-access chain such as CP-OFDM, SC-FDMA, OTFS, filtered OFDM, or SCMA-over-OFDM.
\end{enumerate}

The paper makes five contributions. First, it formulates the two paradigm transitions from design intelligence to runtime intelligence and defines the change scale through explicit invariants. Second, it presents a generate, execute, verify, select, and deploy architecture with hard gates, versioned artifacts, monitoring, and fallback. Third, it evaluates three Level-1 parameter agents using coded baseband Monte Carlo and transparent engineering models. Fourth, it demonstrates two-timescale Level-2 receiver adaptation with actual complex-valued iterative solves and common-random-number MIMO experiments. Fifth, it demonstrates Level-3 waveform adaptation across five representative regimes and continuous sweeps, while explicitly separating generated-baseband evidence from analytical BLER, latency, complexity, and power abstractions.

The scope is intentionally bounded. The current agent chooses and configures human-registered communication tools; it does not invent a new coding theorem, emit production C/CUDA/RTL, or reconfigure an SDR or FPGA. The contribution is a falsifiable minimum viable loop that makes the changing object, decision evidence, and safety boundary explicit.

\section{Related Work and Positioning}
\label{sec:related}

The proposed fabric intersects four research directions. First, cognitive radio and conventional link adaptation sense the environment and choose operating parameters \cite{haykin2005cognitive,3gpp38211}. This is closest to Level~1, although the proposed state also includes task and hardware constraints. Second, AI-native air-interface research studies learned components and end-to-end transceivers \cite{hoydis2021ainative,3gpp38843}. Those methods can learn mappings that are difficult to derive manually, but their principal artifact is usually a trained model. The fabric instead operates on explicit parameters and executable algorithmic candidates, which can be inspected, gated, and mapped to implementation targets.

Third, large-language-model-driven coding agents combine code generation with executable evaluators. AlphaEvolve evolves programs through repeated proposal, execution, scoring, and retention \cite{novikov2025alphaevolve}. AITE transfers this pattern to wireless algorithm discovery and reports communication receivers generated through performance--complexity feedback \cite{aoudia2026aite}. These systems motivate the first transition in this paper: AI changes how algorithms can be developed. The fabric adds the second transition: a deployed agent must decide when a previously validated parameter, solver, or waveform remains appropriate and when a guarded change is necessary.

\begin{figure*}[t]
    \centering
    \includegraphics[width=0.97\textwidth]{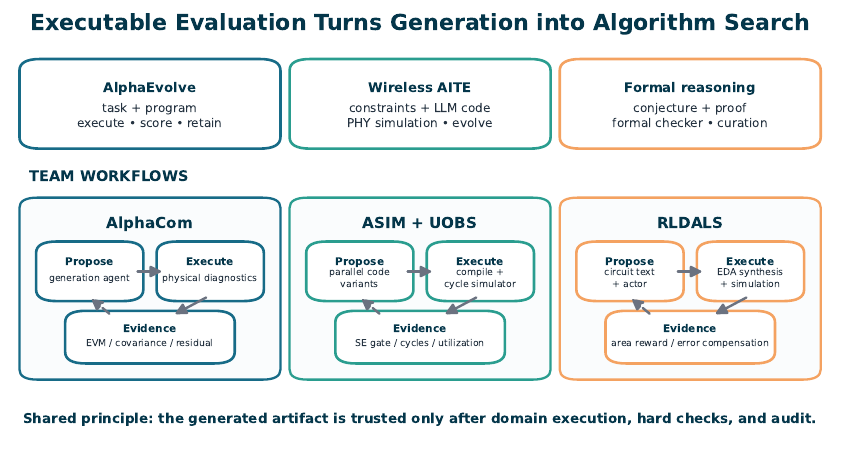}
    \caption{Executable-evaluation workflows related to the proposed fabric. The common pattern is proposal, domain execution, evidence extraction, and revision. AlphaCom uses physical diagnostics, ASIM+UOBS uses numerical and hardware evidence, and RLDALS uses synthesis, simulation, and error compensation.}
    \label{fig:discovery-landscape}
\end{figure*}

Our own design-time work follows the same separation between generative authority and evaluative authority. AlphaCom couples a generation agent with an evaluation agent and returns interpretable PHY diagnostics such as EVM, post-equalization noise covariance, and residual interference rather than only a scalar score \cite{liu2026alphacom}. The ASIM+UOBS prototype compiles generated communication-operator variants, reconstructs numerical behavior, checks spectral-efficiency gates, and measures cycle and utilization evidence; in its Block-Richardson case, accepted scheduling refinements reduce finish cycles from 2550 to 1893, a 25.76\% reduction. RLDALS encodes circuit structure as text, uses an actor to propose legal logic transformations, and places EDA synthesis, Monte Carlo evaluation, and error compensation inside the reinforcement-learning loop \cite{zuo2026rldals}. These examples show why a generated artifact should be treated as a candidate rather than as a result: domain execution and hard validation remain the source of trust.

The same principle appears outside communications. In formal mathematics, cheap proposal generation increases the importance of machine-checkable proof, provenance, interpretation, and human curation. This analogy is useful but limited: a radio action additionally changes a live physical interface, consumes shared spectrum, and can affect a remote endpoint. Runtime communication adaptation therefore requires capability agreement, activation control, monitoring, and rollback in addition to an evaluator.

Fourth, the Level-3 candidate set builds on established waveform and multiple-access research. OTFS maps symbols in the delay--Doppler domain and is motivated by strongly time-varying channels \cite{hadani2017otfs}. Filtered OFDM improves subband spectral localization at the cost of filtering overhead \cite{zhang2015fofdm}. SCMA is a sparse code-domain nonorthogonal multiple-access method rather than a standalone pulse waveform \cite{nikopour2013scma}; accordingly, the present candidate is labeled SCMA-over-OFDM. The paper does not claim a new instance of these algorithms. It studies whether a constrained runtime agent can select among them using a common evidence interface.

\section{Agent-Native Metamorphic Architecture}
\label{sec:architecture}

\begin{figure*}[t]
    \centering
    \includegraphics[width=0.96\textwidth]{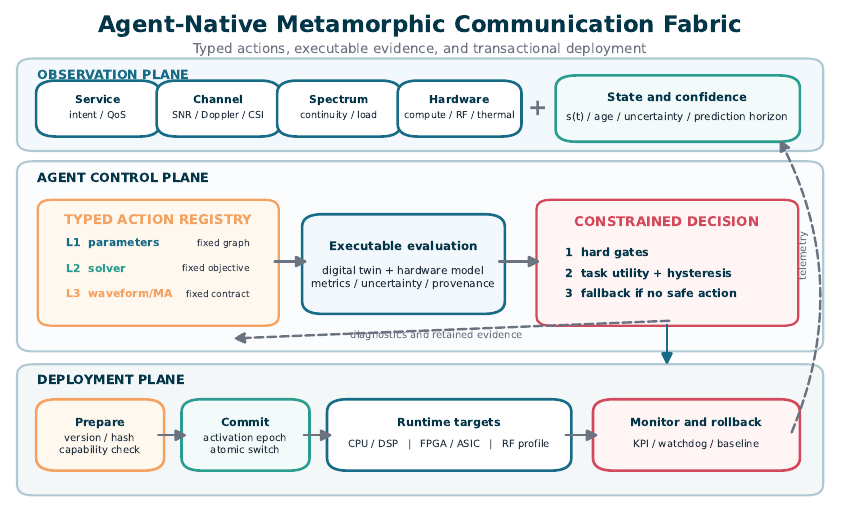}
    \caption{The \fabric organized into observation, agent-control, and deployment planes. Service, channel, spectrum, and hardware telemetry form a confidence-tagged state. A typed Level-1--Level-3 action is evaluated before hard-gated selection, then installed through a prepare--commit transaction with explicit hardware targets, monitoring, and rollback.}
    \label{fig:architecture}
\end{figure*}

\begin{figure*}[t]
    \centering
    \includegraphics[width=0.97\textwidth]{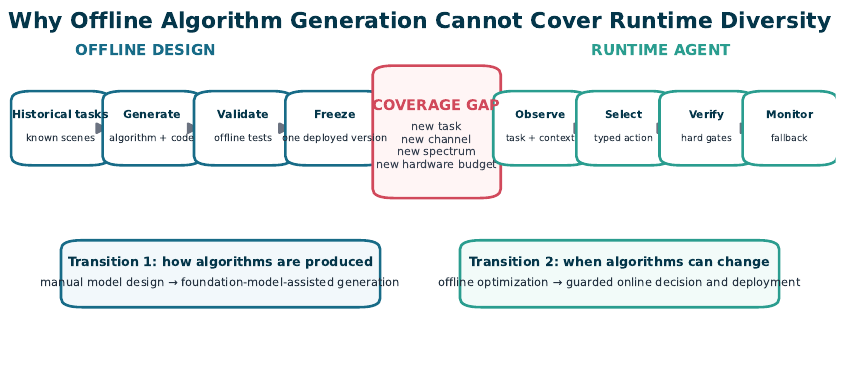}
    \caption{The coverage gap between offline algorithm generation and future runtime combinations. The proposed agent does not discard offline design; it turns validated artifacts into a typed runtime library and re-evaluates their suitability as task, channel, spectrum, load, and hardware budgets change.}
    \label{fig:runtime-gap}
\end{figure*}

At decision epoch $t$, an observer forms
\begin{equation}
\mathbf{s}_t=(\mathbf{q}_t,\mathbf{h}_t,\mathbf{z}_t,\mathbf{c}_t,\mathbf{e}_t),
\label{eq:state}
\end{equation}
where $\mathbf{q}_t$ describes service intent and traffic, $\mathbf{h}_t$ is channel and mobility state, $\mathbf{z}_t$ describes spectrum and offered load, $\mathbf{c}_t$ captures receiver, baseband, and RF capabilities, and $\mathbf{e}_t$ denotes energy and thermal budgets. State uncertainty is handled through conservative gates and fallback rather than by assuming perfect context.

For level $\ell\in\{1,2,3\}$, the agent chooses an action $a\in\mathcal{A}_{\ell}(\mathbf{s}_t)$. Every candidate produces an executable or calculable metric vector
\begin{equation}
\mathbf{m}(a,\mathbf{s}_t)=[R,\eta,\epsilon,L,E,P,C,S],
\end{equation}
containing throughput, effective spectral efficiency, reliability loss, latency, energy, power, complexity, and spectral-containment metrics. A feasible action must satisfy
\begin{equation}
g_j(a,\mathbf{s}_t)=1,\qquad j\in\mathcal{G}_{\ell},
\label{eq:gates}
\end{equation}
where the gate set may include BER/BLER, EVM, residual, divergence, ACLR, PAPR, latency, power, complexity, capability, and hysteresis constraints. Selection occurs only inside the feasible set
\begin{align}
\mathcal{F}_{\ell}(\mathbf{s}_t)&=\{a\in\mathcal{A}_{\ell}(\mathbf{s}_t):g_j=1,\ \forall j\},\\
a_t^*&=\arg\max_{a\in\mathcal{F}_{\ell}(\mathbf{s}_t)}U(\mathbf{m}(a,\mathbf{s}_t);\mathbf{w}_t).
\label{eq:select}
\end{align}
If $\mathcal{F}_{\ell}$ is empty, the agent retains or returns to a registered minimum-risk baseline. This order prevents a weighted utility from trading away a hard reliability or spectral constraint.

The action space expands with level:
\begin{align}
\mathcal{A}_1&=\{\boldsymbol{\theta}\},\\
\mathcal{A}_2&=\{(r,T_{\max},\omega,\mathbf{M},\tau,q)\},\\
\mathcal{A}_3&=\{(w,m,\boldsymbol{\theta}_w)\},
\label{eq:levels}
\end{align}
where $\boldsymbol{\theta}$ is a fixed-topology parameter vector, $r$ is a receiver solver, $T_{\max}$ is an iteration cap, $\omega$ is a relaxation factor, $\mathbf{M}$ is a preconditioner, $\tau$ is an early-stop threshold, $q$ is numerical precision, $w$ is a waveform, and $m$ is an optional multiple-access realization.

The selected action is not passed to the data plane as an unstructured model response. It is serialized as a versioned deployment manifest
\begin{equation}
\mathcal{V}_t=\{\mathrm{id},\ell,h_{\mathrm{impl}},\mathcal{I},\mathbf{r},t_{\mathrm{act}},
\mathrm{id}_{\mathrm{fb}}\},
\label{eq:manifest}
\end{equation}
where $h_{\mathrm{impl}}$ is an implementation hash, $\mathcal{I}$ is the typed input/output interface, $\mathbf{r}$ is the declared compute, memory, RF, and thermal resource envelope, $t_{\mathrm{act}}$ is a future activation epoch, and $\mathrm{id}_{\mathrm{fb}}$ identifies a verified fallback. This representation makes the agent decision auditable and prevents a semantically valid algorithm from being deployed through an incompatible binary, bitstream, numerology, or endpoint interface.

\begin{table*}[t]
\caption{Three levels of runtime communication adaptation.}
\label{tab:levels}
\centering
\footnotesize
\resizebox{\textwidth}{!}{%
\begin{tabular}{lllll}
\toprule
\textbf{Level} & \textbf{Invariant} & \textbf{Agent action} & \textbf{Artifact} & \textbf{Evidence in this paper} \\
\midrule
1: Parameter & Protocol, waveform, algorithm topology & MCS, pilots, regularization, precision & Configuration vector & Coded baseband Monte Carlo and engineering models \\
2: Algorithm & Protocol, waveform, linear-MMSE objective & Solver family, iterations, preconditioner, fallback & Receiver-solver configuration & Complex MIMO Monte Carlo and actual iterative solves \\
3: Waveform/multiple access & Service contract, regulation, safety interface & Signal representation and resource-overload chain & Waveform-chain manifest & Generated baseband samples plus transparent link/RF abstractions \\
\bottomrule
\end{tabular}}
\end{table*}

The levels should not be interpreted as three unrelated optimizers. They are a governance hierarchy. Level~1 can run at a conventional adaptation interval because it selects approved parameters. Level~2 uses a slow timescale to choose a default solver and a fast per-channel-block guard to upgrade unsafe approximate solutions. Level~3 changes the signal representation seen by both endpoints and therefore requires capability agreement, versioning, activation epochs, hysteresis, and rollback. A larger change surface requires stronger evidence and slower deployment.

\subsection{Transactional Hardware Execution Model}

The control plane is intentionally separated from the sample-rate data plane. A foundation model or search process may run asynchronously on a management CPU, but the real-time path executes only signed candidates already present in the typed registry. Deployment follows four phases. \emph{Prepare} checks the manifest hash, interface, memory footprint, worst-case execution budget, endpoint capability, and fallback availability. \emph{Shadow evaluation} runs the candidate on a digital twin, recorded samples, or duplicated live blocks without granting transmit authority. \emph{Commit} performs an atomic switch at a frame, slot, or negotiated activation boundary. \emph{Monitor} compares post-commit telemetry with the admission envelope and triggers rollback if a watchdog, reliability, spectral, or deadline gate fails.

\begin{table*}[t]
\caption{Hardware realization and governance across the three adaptation levels. Timescales are relative architectural classes rather than prescribed standard values.}
\label{tab:hardware-levels}
\centering
\footnotesize
\resizebox{\textwidth}{!}{%
\begin{tabular}{llllll}
\toprule
\textbf{Level} & \textbf{Control timescale} & \textbf{Mutable hardware object} & \textbf{Required data-plane support} & \textbf{Safe activation boundary} & \textbf{Fallback} \\
\midrule
L1 & Link-adaptation interval & Register/LUT configuration vector & Shadow registers, parameterized mapper, pilots, EQ, precision modes & Slot or frame boundary & Last-known-good parameter bank \\
L2 & Scenario update plus per-block guard & Solver ID, iteration cap, precision, preconditioner & Shared matrix memory, MAC/reduction fabric, solver microcode, exact-solver path & Channel-block boundary after state capture & Direct MMSE or certified solver \\
L3 & Session/episode and negotiated event & Versioned waveform/access processing graph and RF profile & Preloaded kernels or FPGA overlays, dual coefficient banks, endpoint capability exchange & Agreed future frame after pipeline drain & Prior waveform chain and RF profile \\
\bottomrule
\end{tabular}}
\end{table*}

This hierarchy supports several implementation styles. A CPU/GPU/SDR prototype can use shared libraries and process isolation; an FPGA can use a static radio shell with parameter registers, preloaded accelerator kernels, or a signed partial-reconfiguration region; and an ASIC can expose a finite family of engines through microcode and clock/power gating. None of these targets should execute arbitrary newly generated code in the real-time path. The agent selects among implementation artifacts whose timing closure, fixed-point behavior, memory layout, and RF interface have already been characterized. Online intelligence therefore changes the admitted artifact and its configuration, while conventional hardware verification remains responsible for the artifact itself.

\subsection{Role of Foundation Models}

The architecture separates proposal generation from deployment authority. A foundation model can structure a task, retrieve a candidate algorithm, generate code, or diagnose a failed evaluation. It cannot directly bypass the typed candidate registry or the gates in (\ref{eq:gates}). The present experiments use deterministic candidate enumeration and task-weighted selection so that every numerical result is reproducible without a remote model. This design choice isolates the central systems question: whether an agentic closed loop can safely adapt the communication object online. Replacing the proposer by an LLM is compatible with the architecture but is not required for the results reported here.

\section{Evaluation Methodology}
\label{sec:method}

All experiments use frozen seeds and machine-readable JSON artifacts. Level~1 uses master seed 20260811 and averages four evaluations per candidate; each evaluation uses 500 coded blocks and 1200 PAPR blocks. Level~2 uses seed 20260818 and evaluates every solver on identical channel matrices, noise, and transmitted symbols. Each scenario contains 700 independent channel blocks. The Level-2 SNR sweep uses 500 independent blocks per SNR point. Level~3 also uses seed 20260818; its waveform engine generates oversampled complex baseband samples for PAPR and adjacent-band leakage proxies, while BLER, processing delay, complexity, PA efficiency, and total power are transparent analytical abstractions.

The reported results are simulation evidence, not hardware measurements. In Level~2, solver quality, residuals, BER, EVM, condition number, and convergence are produced by actual complex matrix solves. Energy and latency are calculated by a common 500-MHz, 64-lane analytical hardware proxy; channel estimation, Gram-matrix formation, and matched filtering are excluded because they are common to every candidate. In Level~3, PAPR and leakage use generated samples, but the remaining link and hardware metrics are model-based. These boundaries are retained in every interpretation below.

\section{Level 1: Parameter Adaptation}
\label{sec:level1}

\begin{figure*}[t]
    \centering
    \includegraphics[width=0.97\textwidth]{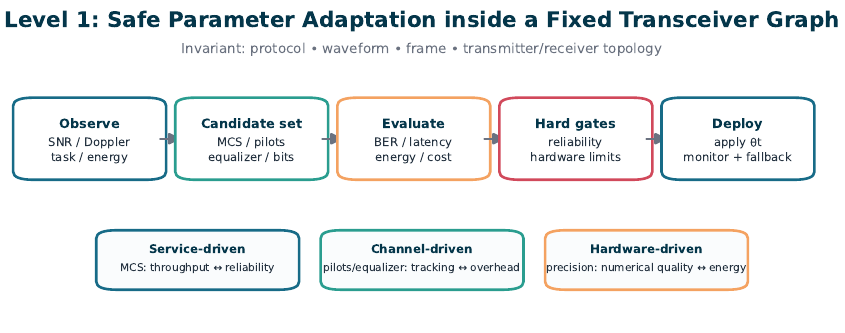}
    \caption{Level-1 control translated from the presentation workflow. A fixed transceiver graph exposes approved service-, channel-, and hardware-driven parameter axes. Evaluation and hard gates precede deployment, and an empty feasible set returns the baseline.}
    \label{fig:l1-control}
\end{figure*}

Level~1 keeps the transceiver topology fixed and selects
\begin{equation}
\boldsymbol{\theta}=[M,R_c,N_{\mathrm{CP}},\rho_p,\lambda_{\mathrm{eq}},b_q],
\label{eq:theta}
\end{equation}
where $M$ is QAM order, $R_c$ is code rate, $N_{\mathrm{CP}}$ is cyclic-prefix length, $\rho_p$ is pilot fraction, $\lambda_{\mathrm{eq}}$ controls MMSE regularization, and $b_q$ is complex-baseband quantization precision. Three agents isolate three common operational decisions.

\subsection{State, Robust Gates, and Hardware Mapping}

A Level-1 decision uses a measurement window rather than one instantaneous channel estimate. The observer aggregates SNR, Doppler, CSI age, decoder statistics, queue occupancy, deadline slack, and accelerator counters, and attaches an age and confidence to every field. For an upper-bounded metric such as BER, latency, power, or temperature, a conservative admission test is
\begin{equation}
\widehat m_j(\boldsymbol{\theta})+\beta_j\sigma_j(\boldsymbol{\theta})\leq b_j,
\label{eq:l1confidence}
\end{equation}
with the inequality reversed for lower-bounded rate or spectral efficiency. Here $\sigma_j$ captures estimator variation and model mismatch, and $\beta_j$ sets the operating margin. Thus the utility function ranks only candidates whose confidence-adjusted metrics pass; it cannot exchange a small expected energy gain for an uncertain reliability violation.

On hardware, the candidate vector maps naturally to a shadow configuration bank. $M$ and $R_c$ select mapper and coding-rate entries, $N_{\mathrm{CP}}$ and $\rho_p$ program framing and pilot address generators, $\lambda_{\mathrm{eq}}$ selects regularization coefficients, and $b_q$ selects a preverified fixed-point mode. The active bank is not edited field by field. The controller writes a complete inactive bank, checks cross-field consistency, and atomically swaps banks at a frame boundary; a watchdog can restore the previous bank without reconstructing the configuration. Multi-precision selection requires a datapath synthesized for the admitted widths or separate prebuilt kernels---changing $b_q$ does not imply arbitrary runtime resynthesis.

\begin{algorithm}[t]
\caption{Level-1 Constrained Online Parameter Adaptation}
\label{alg:l1}
\begin{algorithmic}[1]
\REQUIRE Runtime state $\mathbf{s}_t$, approved set $\Theta$, baseline $\boldsymbol{\theta}_0$
\ENSURE Safe configuration $\boldsymbol{\theta}_t$
\STATE Observe SNR, Doppler, CSI age, load, energy, and hardware capability
\STATE Retrieve candidates compatible with the fixed waveform and topology
\FORALL{$\boldsymbol{\theta}\in\Theta$}
    \STATE Estimate BER, latency, complexity, energy, and confidence under $\mathbf{s}_t$
    \STATE Discard $\boldsymbol{\theta}$ if any confidence-adjusted reliability or hardware gate fails
    \STATE Compute task utility $J(\boldsymbol{\theta}\mid\mathbf{s}_t)$
\ENDFOR
\STATE Select $\boldsymbol{\theta}_t=\arg\max J$; use $\boldsymbol{\theta}_0$ if the feasible set is empty
\STATE Program and verify an inactive configuration bank
\STATE Atomically activate $\boldsymbol{\theta}_t$ at the agreed boundary
\STATE Monitor online metrics and restore $\boldsymbol{\theta}_0$ on gate violation
\end{algorithmic}
\end{algorithm}

\begin{table}[t]
\caption{Level-1 Registered Candidate Axes}
\label{tab:l1space}
\centering
\footnotesize
\resizebox{\columnwidth}{!}{%
\begin{tabular}{lll}
\toprule
\textbf{Agent} & \textbf{Variable candidates} & \textbf{Count} \\
\midrule
MCS & $M\in\{4,16,64\}$, $R_c\in\{1/2,3/4\}$ & 6 \\
Tracking & $\rho_p\in\{0.08,0.16\}$, EQ $\in\{$MMSE, robust-MMSE$\}$ & 4 \\
Precision & $b_q\in\{8,12\}$ bit & 2 \\
\bottomrule
\end{tabular}
}
\end{table}

\begin{table*}[t]
\caption{Level-1 parameter agents. Relative changes use the fixed-parameter baseline in the same scenario.}
\label{tab:l1}
\centering
\footnotesize
\resizebox{\textwidth}{!}{%
\begin{tabular}{llllrrrrl}
\toprule
\textbf{Agent} & \textbf{Fixed topology} & \textbf{Selected parameters} & \textbf{BER} & \textbf{Rate} & \textbf{Latency} & \textbf{Energy} & \textbf{Main change vs. baseline} & \textbf{Interpretation} \\
\midrule
MCS & CP-OFDM, robust MMSE, 8 bit & 16-QAM, $R_c=3/4$ & $7.75\!\times\!10^{-3}$ & 42.16 Mb/s & 796.05 ms & 39.70 nJ/bit & Rate $+200\%$, latency $-66.7\%$, energy $-67.4\%$ & Use available reliability margin \\
Tracking & DFT-s-OFDM, QPSK, $R_c=3/4$, 8 bit & Pilot 16\%, MMSE & $4.668\!\times\!10^{-2}$ & 3.78 Mb/s & 208.20 ms & 178.56 nJ/bit & BER $-36.6\%$; rate $-8.7\%$, latency $+9.5\%$, energy $+10.2\%$ & Spend resources on stale-CSI robustness \\
Precision & CP-OFDM, 16-QAM, $R_c=3/4$, robust MMSE & 8-bit baseband & $1.873\!\times\!10^{-3}$ & 18.40 Mb/s & 57.14 ms & 13.44 nJ/bit & Same rate/latency, energy $-8.0\%$ & Reduce precision inside reliability gate \\
\bottomrule
\end{tabular}}
\end{table*}

The MCS agent converts excess reliability margin into rate and energy efficiency. The tracking agent illustrates the opposite decision: higher pilot overhead and a changed equalizer setting reduce BER but impose measurable rate, latency, and energy penalties. The precision agent retains link performance while selecting the lowest approved precision that passes the BER gate. These results are operationally useful, but they do not constitute algorithm generation because the computation graph remains fixed.

The minimum hardware controller for this level is a finite-state machine with telemetry capture, candidate-table lookup, gate evaluation, shadow-bank programming, and rollback states. Local choices such as equalizer regularization or precision can be receiver-only. Changes to MCS, pilot pattern, or cyclic prefix are dual-endpoint actions and must be signaled before the activation epoch. Their control signaling, bank-switch latency, and transition guard interval should be charged to the candidate when adaptation is frequent; otherwise a rapidly oscillating policy could appear efficient in steady-state metrics while wasting resources during reconfiguration.

\section{Level 2: Receiver Algorithm Adaptation}
\label{sec:level2}

\begin{figure*}[t]
    \centering
    \includegraphics[width=0.97\textwidth]{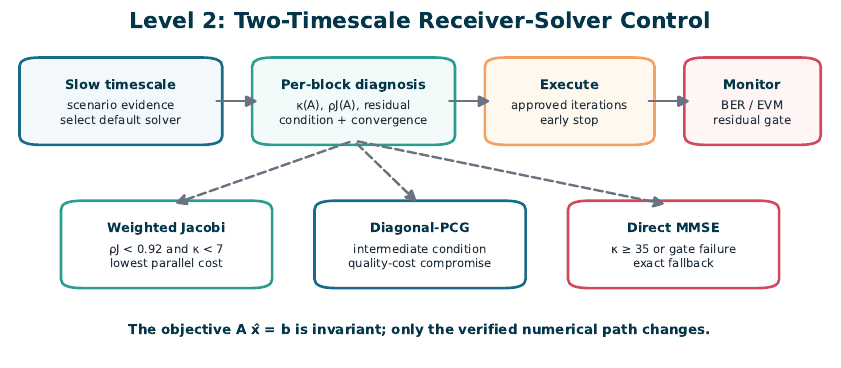}
    \caption{Level-2 two-timescale control. A slow decision selects a default solver from scenario evidence. A fast guard observes each channel block and upgrades the numerical path when conditioning, spectral-radius, residual, EVM, or convergence evidence becomes unsafe.}
    \label{fig:l2-control}
\end{figure*}

Level~2 fixes the protocol, waveform, number of streams, QAM order, and linear-MMSE objective. For
\begin{equation}
\mathbf{y}=\mathbf{H}\mathbf{x}+\mathbf{n},
\end{equation}
the receiver solves
\begin{equation}
\mathbf{A}\hat{\mathbf{x}}=\mathbf{b},\qquad
\mathbf{A}=\hat{\mathbf{H}}^{H}\hat{\mathbf{H}}+\sigma_n^2\mathbf{I},\quad
\mathbf{b}=\hat{\mathbf{H}}^{H}\mathbf{y}.
\label{eq:mmse}
\end{equation}
Direct MMSE, weighted Jacobi, and diagonal-PCG solve the same system. Their difference is the numerical path, convergence region, approximation error, and hardware cost.

Direct MMSE uses a Hermitian positive-definite factorization and triangular solves. Weighted Jacobi uses $\mathbf{D}=\diag(\mathbf{A})$ and
\begin{equation}
\mathbf{x}^{(k+1)}=(1-\omega)\mathbf{x}^{(k)}+
\omega\mathbf{D}^{-1}\left[\mathbf{b}-(\mathbf{A}-\mathbf{D})\mathbf{x}^{(k)}\right].
\label{eq:jacobi}
\end{equation}
It is attractive when the iteration matrix has spectral radius below one and the Gram matrix is close to diagonally dominant. Diagonal-PCG uses $\mathbf{M}=\diag(\mathbf{A})$ as a preconditioner and can cover the intermediate region in which Jacobi is unsafe but a small number of Krylov iterations remains sufficient.

\subsection{Two-Timescale Guard and Solver Hardware}

The common channel-estimation front end forms $\mathbf{A}$ and $\mathbf{b}$ once and writes them to a solver-visible matrix buffer. A slow controller selects the nominal solver configuration from scenario statistics, while a per-block guard observes conditioning and convergence evidence before and during execution. Weighted Jacobi requires matrix-vector products and elementwise scaling; PCG additionally requires dot-product reductions, vector recurrences, and a preconditioner application; direct MMSE requires a factorization engine and triangular solves. A practical architecture can share the complex multiply--accumulate array and SRAM banks across all three paths while retaining separate reduction and factorization controllers.

For a solver configuration $r$, the analytical hardware proxy can be written structurally as
\begin{align}
L_r &\approx \frac{1}{f_{\mathrm{clk}}}\left[
\left\lceil\frac{N_{\mathrm{MAC},r}}{P_{\mathrm{MAC}}}\right\rceil+
N_{\mathrm{red},r}d_{\mathrm{red}}\right.\notag\\
&\qquad\left.+
\left\lceil\frac{N_{\mathrm{mem},r}}{B_{\mathrm{mem}}}\right\rceil+d_{\mathrm{ctrl},r}\right],
\label{eq:l2latency}\\
E_r &\approx N_{\mathrm{MAC},r}e_{\mathrm{MAC}}(q)+
N_{\mathrm{mem},r}e_{\mathrm{mem}}(q)\notag\\
&\qquad+
N_{\mathrm{red},r}e_{\mathrm{red}}+E_{\mathrm{ctrl},r}.
\label{eq:l2energy}
\end{align}
The reported 500-MHz, 64-lane results instantiate this operation-count model with a common parallelism assumption. Equations~(\ref{eq:l2latency})--(\ref{eq:l2energy}) expose what must be replaced by synthesis or measurements: memory-bank conflicts, pipeline fill/drain, reduction depth, precision-dependent MAC energy, and control overhead.

The experiments compute $\kappa(\mathbf{A})$ and $\rho_J(\mathbf{A})$ exactly to characterize the regions. A deployed per-block guard need not perform an expensive full eigendecomposition. It can use a short power/Lanczos estimate, Gershgorin bounds, diagonal-dominance statistics, residual contraction $\|\mathbf{r}^{(k)}\|/\|\mathbf{r}^{(k-1)}\|$, or a learned but conservatively calibrated condition classifier. Regardless of the proxy, the running residual and EVM watchdog remain authoritative. A failed early-stop or deadline check cancels the approximate path and dispatches the same buffered $\mathbf{A},\mathbf{b}$ to the direct solver, avoiding a second channel-estimation pass.

\begin{table*}[t]
\caption{Complete registered Level-2 receiver-solver configuration library. All configurations solve (\ref{eq:mmse}); the differences are the numerical path and iteration budget.}
\label{tab:l2library}
\centering
\footnotesize
\begin{tabular}{llllll}
\toprule
\textbf{ID} & \textbf{Solver} & $T_{\max}$ & $\omega$ & $\tau$ & \textbf{Role} \\
\midrule
J2 & Weighted Jacobi & 2 & 0.90 & $10^{-3}$ & Minimum-cost probe \\
J3 & Weighted Jacobi & 3 & 0.90 & $10^{-3}$ & Favorable-channel default \\
J4 & Weighted Jacobi & 4 & 0.90 & $10^{-3}$ & Higher-accuracy Jacobi \\
J6 & Weighted Jacobi & 6 & 0.82 & $10^{-3}$ & Conservative relaxation \\
P2 & Diagonal-PCG & 2 & -- & $10^{-3}$ & Low-order Krylov probe \\
P3 & Diagonal-PCG & 3 & -- & $10^{-3}$ & SNR-sweep intermediate action \\
P5 & Diagonal-PCG & 5 & -- & $10^{-3}$ & Moderate-condition default \\
D & Direct MMSE & factorization & -- & exact & Reference and final fallback \\
\bottomrule
\end{tabular}
\end{table*}

The agent chooses
\begin{equation}
a=(r,T_{\max},\omega,\mathbf{M},\tau,q)
\end{equation}
to minimize normalized power, latency, complexity, and numerical error subject to BER, EVM, residual-divergence, and P95-latency gates. The slow-timescale decision selects a default solver from scenario statistics. The fast guard recomputes the Gram condition number $\kappa(\mathbf{A})$ and Jacobi spectral radius for every channel block. The current empirical safety region keeps Jacobi only when $\rho<0.92$ and $\kappa<7$, upgrades to PCG in the intermediate region, and returns to direct MMSE when $\kappa\geq35$ or the residual gate fails. The thresholds define the present test policy, not universal channel constants.

\begin{algorithm}[t]
\caption{Level-2 Two-Timescale Receiver-Solver Selection}
\label{alg:l2}
\begin{algorithmic}[1]
\REQUIRE Scenario statistics $\mathcal{S}$, block matrix $\mathbf{A}$, vector $\mathbf{b}$
\ENSURE Estimate $\hat{\mathbf{x}}$, solver ID, residual, and fallback reason
\STATE At the slow timescale, select a default feasible solver from Monte Carlo evidence
\STATE Estimate conditioning and Jacobi stability for the current block
\IF{$\rho_J<0.92$ and $\kappa(\mathbf{A})<7$}
    \STATE choose weighted Jacobi
\ELSE
    \STATE choose diagonal-PCG
\ENDIF
\IF{$\kappa(\mathbf{A})\geq35$}
    \STATE upgrade immediately to Direct MMSE
\ENDIF
\STATE Run the selected solver with approved iterations and early stopping
\IF{residual, EVM, convergence, or deadline gate fails}
    \STATE fall back to Direct MMSE
\ENDIF
\STATE Log solver, iterations, residual trajectory, cycles, stalls, saturation, and fallback reason
\end{algorithmic}
\end{algorithm}

\begin{table*}[t]
\caption{Level-2 scenario results. Energy and latency reductions are relative to direct MMSE and are predictions of the common analytical hardware proxy.}
\label{tab:l2}
\centering
\footnotesize
\resizebox{\textwidth}{!}{%
\begin{tabular}{lrrrrlrrrr}
\toprule
\textbf{Scenario} & $N_r/K$ & \textbf{SNR} & $\mathrm{median}\,\kappa$ & $\rho_{J,90}$ & \textbf{Default action} & \textbf{BER} & \textbf{EVM} & \textbf{Energy red.} & \textbf{Latency red.} \\
\midrule
Favorable short-range massive MIMO & 48/8 & 22 dB & 3.85 & 0.81 & Jacobi-3, $\omega=0.9$ & 0 & 10.76\% & 67.1\% & 73.3\% \\
Moderately correlated mobile MIMO & 16/8 & 15 dB & 22.45 & 2.17 & Diagonal-PCG-5 & $2.364\!\times\!10^{-2}$ & 22.56\% & 10.0\% & 22.5\% \\
Harsh correlated edge MIMO & 12/10 & 10 dB & 61.91 & 5.00 & Direct MMSE & $9.043\!\times\!10^{-2}$ & 59.53\% & 0 & 0 \\
\bottomrule
\end{tabular}}
\end{table*}

\begin{figure*}[t]
    \centering
    \includegraphics[width=0.97\textwidth]{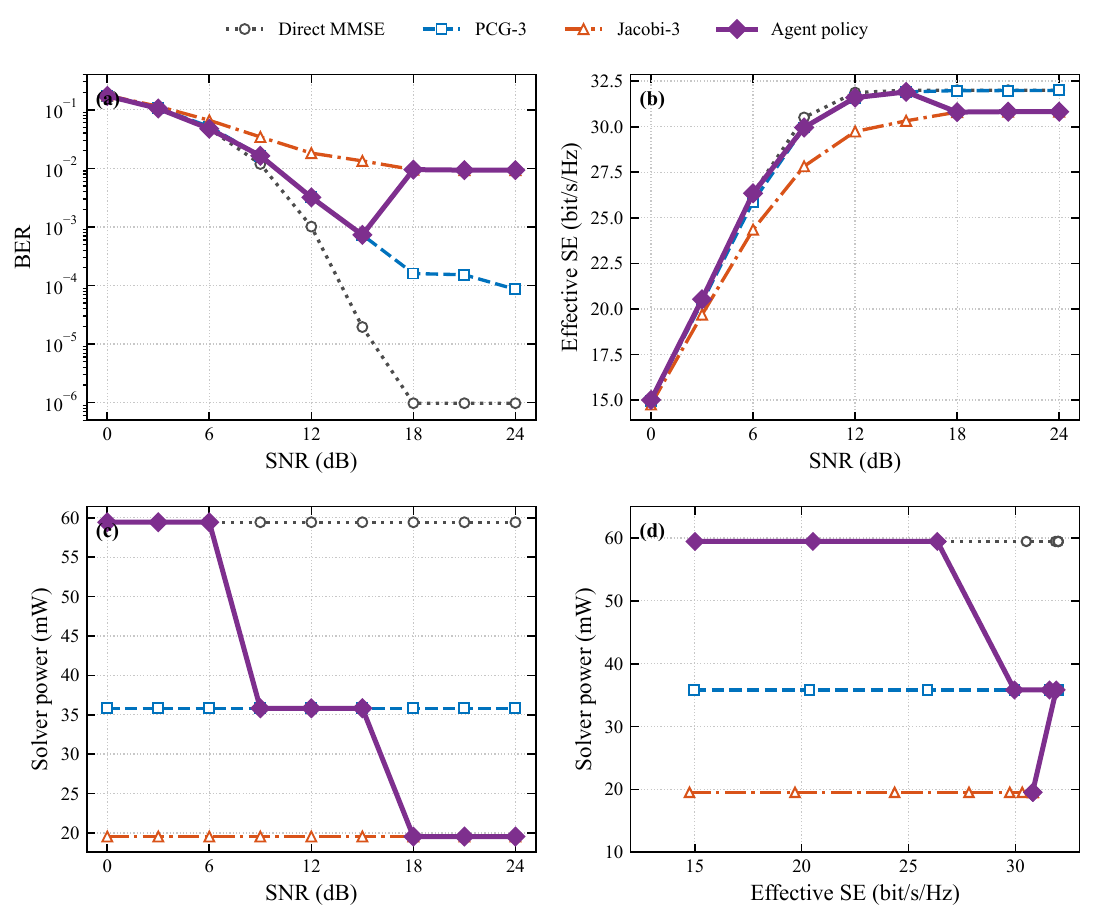}
    \caption{Level-2 receiver adaptation. The panels show BER versus SNR, effective spectral efficiency versus SNR, analytical solver power versus SNR, and the spectral-efficiency--power tradeoff. The thick envelope is the agent choice after quality gates.}
    \label{fig:l2}
\end{figure*}

Table~\ref{tab:l2} shows a stable three-region policy. In favorable channels, three Jacobi iterations pass the EVM and reliability gates while eliminating factorization. In the intermediate scenario, low-order Jacobi diverges and PCG-5 becomes the lowest-cost feasible solver. In the harsh scenario, PCG-5 passes the loose feasibility gate but direct MMSE is retained because reliability has dominant weight and the direct solution has lower BER, EVM, and residual.

The per-block guard makes the result online rather than a three-scenario lookup. In the favorable scenario, 1.57\% of blocks are upgraded from Jacobi to PCG. In the intermediate scenario, 9.29\% are upgraded from PCG to direct MMSE, reducing BER from $2.364\times10^{-2}$ to $2.116\times10^{-2}$ after guarded execution. The harsh scenario remains on direct MMSE for all blocks.

Hardware feasibility depends on fallback capacity, not only average solver cost. If the direct engine is time-multiplexed across users, admission control must reserve enough slack for the measured upgrade probability and its tail, or the fallback itself can miss the deadline. Warm starts may reduce iterations, but the initial vector, residual, precision, and stopping rule must be versioned with the solver configuration. The implementation log therefore records solver ID, actual iteration count, residual trajectory, numerical saturation flags, cycle count, memory stalls, and whether the block completed on the approximate or exact path.

The SNR sweep in Fig.~\ref{fig:l2} provides a second view of the working regions. Under a BER gate of 0.04 and an effective-spectral-efficiency requirement of at least 95\% of direct MMSE, the agent uses direct MMSE at 0--6 dB, PCG-3 at 9--15 dB, and Jacobi-3 at 18--24 dB. The corresponding analytical compute powers are 59.44, 35.82, and 19.53 mW at 10 million vectors/s. Thus PCG-3 and Jacobi-3 reduce modeled solver power by 39.7\% and 67.1\%, respectively. At low SNR, direct MMSE is a minimum-risk fallback; it is not claimed to satisfy an absolute BER target that no candidate can meet.

\section{Level 3: Waveform and Multiple-Access Adaptation}
\label{sec:level3}

\begin{figure*}[t]
    \centering
    \includegraphics[width=0.97\textwidth]{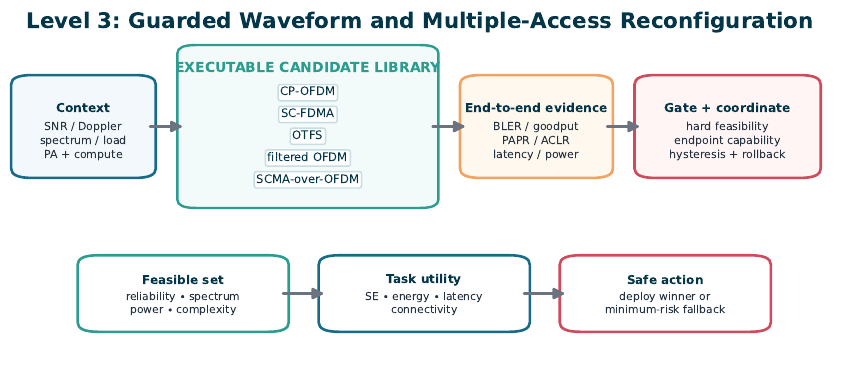}
    \caption{Level-3 waveform and multiple-access control. The agent evaluates executable baseband chains under common evidence and gates, maximizes utility only inside the feasible set, and coordinates endpoint capability, hysteresis, and rollback before changing the transmitted representation.}
    \label{fig:l3-control}
\end{figure*}

Level~3 changes the physical signal representation while retaining the service contract, regulatory limits, capability exchange, and fallback behavior. The agent observes
\begin{equation}
\begin{aligned}
\mathbf{s}^{(3)}=\{&\gamma,\nu,\tau_d/T_{\mathrm{CP}},c_f,\rho_L,\\
&P_{\mathrm{RF}},P_{\mathrm{BB}},R_{\mathrm{ACLR}},D_{\max},
\epsilon_{\mathrm{BLER}},\mathbf{w}\},
\end{aligned}
\label{eq:l3state}
\end{equation}
where $\gamma$ is SNR, $\nu$ normalized Doppler, $c_f$ spectrum contiguity, $\rho_L$ offered load normalized by orthogonal resources, and the remaining terms specify RF/baseband budgets and task gates.

\subsection{Waveform Manifest and Coordinated Switching}

A Level-3 action must describe an executable chain, not only a waveform name. Its manifest instantiates
\begin{equation}
\begin{aligned}
\mathcal{V}_w=\{&w,\mu,N_{\mathrm{FFT}},N_{\mathrm{CP}},\mathbf{p},h_{\mathrm{coef}},\\
&r_{\mathrm{RX}},\pi_{\mathrm{state}},\pi_{\mathrm{RF}},t_{\mathrm{act}},w_{\mathrm{fb}}\},
\end{aligned}
\label{eq:l3manifest}
\end{equation}
where $\mu$ is numerology, $\mathbf{p}$ is the resource map, $h_{\mathrm{coef}}$ identifies filter, spreading, or codebook coefficients, $r_{\mathrm{RX}}$ identifies the matching receiver, and $\pi_{\mathrm{state}}$ specifies whether channel estimates, synchronization loops, and HARQ soft state are translated, version-tagged, or flushed. The RF profile $\pi_{\mathrm{RF}}$ declares sample rate, occupied band, PA operating point, and spectral-mask calibration. Both endpoints acknowledge the same manifest hash and activation epoch before transmit representation changes.

The gate must include transition cost as well as steady-state performance. For candidate $w$, define
\begin{align}
D_{\mathrm{sw},w}&=D_{\mathrm{drain}}+D_{\mathrm{load}}+D_{\mathrm{sync}}+D_{\mathrm{RF}},\\
E_{\mathrm{sw},w}&=E_{\mathrm{state}}+E_{\mathrm{memory}}+E_{\mathrm{RF}},
\label{eq:l3switchcost}
\end{align}
covering pipeline drain, kernel/coefficient loading, reacquisition, RF settling, and state movement. These terms are amortized over the expected residence time of the new action. A switch is suppressed unless the predicted utility improvement exceeds its transition cost and a hysteresis margin; this prevents Doppler, spectrum, or load estimates near a boundary from causing repeated waveform oscillation.

\begin{table*}[t]
\caption{Complete Level-3 executable waveform and multiple-access library. $\mathbf{F}_N$ is a unitary DFT matrix, $\mathbf{P}$ maps active resources, $\mathbf{C}$ inserts the cyclic prefix, $\mathbf{H}_f$ is the implemented subband filter, and $\mathbf{c}_u$ is a sparse SCMA signature.}
\label{tab:l3tools}
\centering
\footnotesize
\resizebox{\textwidth}{!}{%
\begin{tabular}{lllll}
\toprule
\textbf{Tool} & \textbf{Executable transmit construction} & \textbf{Receiver representation} & \textbf{Principal advantage} & \textbf{Cost or risk} \\
\midrule
CP-OFDM & $\mathbf{x}=\mathbf{C}\mathbf{F}_{N}^{H}\mathbf{P}\mathbf{s}$ & FFT and one-tap subcarrier equalization & Flexible scheduling and mature processing & High PAPR and Doppler-induced ICI \\
SC-FDMA & $\mathbf{x}=\mathbf{C}\mathbf{F}_{N}^{H}\mathbf{P}\mathbf{F}_{M}\mathbf{s}$ & FFT, de-map, IDFT, equalize & Lower PAPR and improved PA efficiency & Less flexible resource allocation \\
OTFS & $\mathbf{X}_{tf}=\mathbf{F}_{\tau}^{H}\mathbf{S}_{dd}\mathbf{F}_{\nu}$; OFDM per slot & 2-D time-frequency/delay-Doppler detection & Stable representation under mobility & Higher detection, storage, and processing cost \\
Filtered OFDM & $\mathbf{x}=\mathbf{H}_{f}\mathbf{x}_{\mathrm{OFDM}}$ (257-tap Kaiser FIR) & Subband filter then OFDM receiver & Improved adjacent-band containment & Filter transient, latency, and multiply cost \\
SCMA-over-OFDM & $x_r=\sum_{u=1}^{U}c_{u,r}s_u$ followed by OFDM & Sparse multiuser detection over resources & Load beyond orthogonal resources & Codebook and message-passing complexity \\
\bottomrule
\end{tabular}}
\end{table*}

The five candidates imply different hardware structures. CP-OFDM and SC-FDMA can share FFT engines, cyclic-prefix buffers, and resource mappers, with SC-FDMA adding DFT spreading. Filtered OFDM adds a coefficient-banked FIR/polyphase stage whose group delay and multiplier occupancy must be scheduled. OTFS can reuse FFT primitives but requires two-dimensional transforms, delay--Doppler storage, and a detector with a larger working set. SCMA-over-OFDM preserves the OFDM modulator but adds codebook lookup and iterative multiuser message passing at the receiver. A practical FPGA/ASIC therefore benefits from a static streaming shell, shared transform and memory primitives, and preverified plug-in engines rather than five independent complete modems.

For SDR and FPGA prototypes, preloading all admitted kernels gives the lowest switching latency but consumes more memory and logic. Partial reconfiguration reduces resident area but introduces bitstream-transfer and configuration delay and must preserve the static timing and I/O contract. For an ASIC, Level-3 freedom is necessarily bounded by the engines fabricated in silicon; the agent can choose microcode, coefficients, power states, and routing among them, but cannot materialize an unanticipated signal path. Analog and mixed-signal resources impose an additional boundary: PA bias, filters, PLLs, and ADC/DAC rates expose only calibrated control ranges, and RF settling plus EVM/ACLR revalidation must complete before the commit is considered successful.

For candidate $w$, the hard gates are
\begin{equation}
\begin{aligned}
\mathrm{BLER}_w&\leq\epsilon_{\mathrm{BLER}}, &
\mathrm{ACLR}_w&\geq R_{\mathrm{ACLR}},\\
D_w&\leq D_{\max}, & P_w&\leq P_{\max}, & C_w&\leq C_{\max}.
\end{aligned}
\label{eq:l3gates}
\end{equation}
Only the feasible set participates in the task-weighted decision
\begin{equation}
\begin{aligned}
w^*=\arg\max_{w\in\mathcal{F}_3}\big(&w_{SE}\bar\eta_w+w_R\bar R_w-w_E\bar E_w\\
&-w_D\bar D_w-w_C\bar C_w+w_S\overline{\mathrm{ACLR}}_w\\
&+w_U\bar U_w\big).
\end{aligned}
\label{eq:l3utility}
\end{equation}
The candidate library is not a set of scene labels. Every tool is evaluated in every scene, candidates that fail a hard gate are removed, and the normalized objective selects among the survivors.

\begin{algorithm}[t]
\caption{Level-3 Safe Waveform and Multiple-Access Adaptation}
\label{alg:l3}
\begin{algorithmic}[1]
\REQUIRE State $\mathbf{s}$, candidate library $\mathcal{W}$, task weights $\mathbf{w}$
\ENSURE Deployable waveform chain $w^*$ or minimum-risk fallback
\STATE Observe SNR, Doppler, delay spread, spectrum contiguity, load, and PA budget
\FORALL{$w\in\mathcal{W}$}
    \STATE Generate executable baseband blocks; measure PAPR and leakage proxy
    \STATE Evaluate BLER, goodput SE, latency, power, and complexity
    \STATE Keep $w$ only if reliability, spectral, power, and complexity gates pass
\ENDFOR
\IF{the feasible set is nonempty}
    \STATE maximize weighted task utility
\ELSE
    \STATE minimize violated gates and risk
\ENDIF
\STATE Charge pipeline, loading, synchronization, RF, and state-transfer costs
\STATE Apply hysteresis and endpoint capability checks before switching
\STATE Build and acknowledge a versioned dual-endpoint manifest
\STATE Pre-stage kernels, coefficients, state policy, RF profile, and fallback
\STATE Commit $w^*$ at the agreed epoch; monitor and roll back on violation
\end{algorithmic}
\end{algorithm}

\begin{table*}[t]
\caption{Level-3 formal operating points from seed 20260818. Power and energy use transparent analytical models; PAPR and leakage use generated baseband samples.}
\label{tab:l3results}
\centering
\footnotesize
\resizebox{\textwidth}{!}{%
\begin{tabular}{llllrrrrl}
\toprule
\textbf{Operating regime} & \textbf{Selected tool} & \textbf{Defining state} & \textbf{Goodput SE} & \textbf{BLER} & \textbf{Power} & \textbf{Energy} & \textbf{Additional evidence} & \textbf{Selection reason} \\
\midrule
Continuous-spectrum downlink & CP-OFDM & Continuous band, broadband & 2.320 & $2.0\!\times\!10^{-4}$ & 1.427 W & 30.8 nJ/bit & -- & Flexible high-rate allocation \\
PA-limited terminal uplink & SC-FDMA & Tight RF budget & 1.820 & $3.78\!\times\!10^{-3}$ & 0.459 W & 25.24 nJ/bit & PAPR$_{99}=8.57$ dB & Higher PA efficiency \\
High-mobility doubly selective & OTFS & $\nu=0.30$ & 1.976 & $7.92\!\times\!10^{-3}$ & 0.935 W & 47.3 nJ/bit & Complexity factor 2.75 & Doppler robustness \\
Fragmented industrial spectrum & f-OFDM & $c_f=0.42$ & 1.889 & $6.16\!\times\!10^{-4}$ & 0.610 W & 32.3 nJ/bit & ACLR proxy 22.66 dB & Adjacent-band gate \\
Dense grant-free access & SCMA-over-OFDM & $\rho_L=1.38$ & 2.645 & $1.06\!\times\!10^{-2}$ & 0.600 W & 22.7 nJ/bit & P95 latency 9.92 ms & Overloaded connectivity \\
\bottomrule
\end{tabular}}
\end{table*}

\subsection{Interpretation of the Selected Waveform Chains}

\textbf{Continuous downlink and PA-limited uplink:} CP-OFDM remains the selected action for the continuous-spectrum broadband downlink because its 2.320 bit/s/Hz goodput SE and mature allocation flexibility dominate the current objective. The agent therefore preserves the conventional waveform rather than changing for its own sake. In the terminal uplink, SC-FDMA is selected because DFT spreading reduces PAPR$_{99}$ from the CP-OFDM region to 8.57 dB. Under the present PA model, this yields 25.24 nJ/bit while keeping BLER at $3.78\times10^{-3}$.

\textbf{High mobility:} at normalized Doppler $\nu=0.30$, the OFDM-derived candidates in the current abstraction lose their reliability and latency gates, while the OTFS delay--Doppler representation remains feasible. OTFS provides 1.976 bit/s/Hz at BLER $7.92\times10^{-3}$, but its processing-complexity factor is 2.75. The selection is thus a context-dependent exchange of compute for mobility robustness, not a universal preference for OTFS.

\textbf{Fragmented spectrum:} with spectrum contiguity $c_f=0.42$ and a 22-dB ACLR gate, only the filtered-OFDM chain remains spectrally feasible. Its generated waveform reaches a 22.66-dB leakage proxy and 1.889 bit/s/Hz goodput SE. The 257-tap Kaiser-window FIR is intentionally explicit in the candidate manifest so that filter delay and multiply cost participate in the decision.

\textbf{Overloaded access:} at offered load $\rho_L=1.38$, orthogonal candidates exceed their supported-load region. SCMA-over-OFDM uses six sparse layers over four-resource groups, then maps the superposed resources through the OFDM chain. It is the only feasible candidate in the formal scene, reaching 2.645 bit/s/Hz aggregate goodput SE, BLER $1.06\times10^{-2}$, and 9.92-ms P95 latency. The higher complexity factor of 4.20 remains visible and becomes decisive when the connectivity requirement is relaxed.

\begin{figure*}[t]
    \centering
    \includegraphics[width=0.97\textwidth]{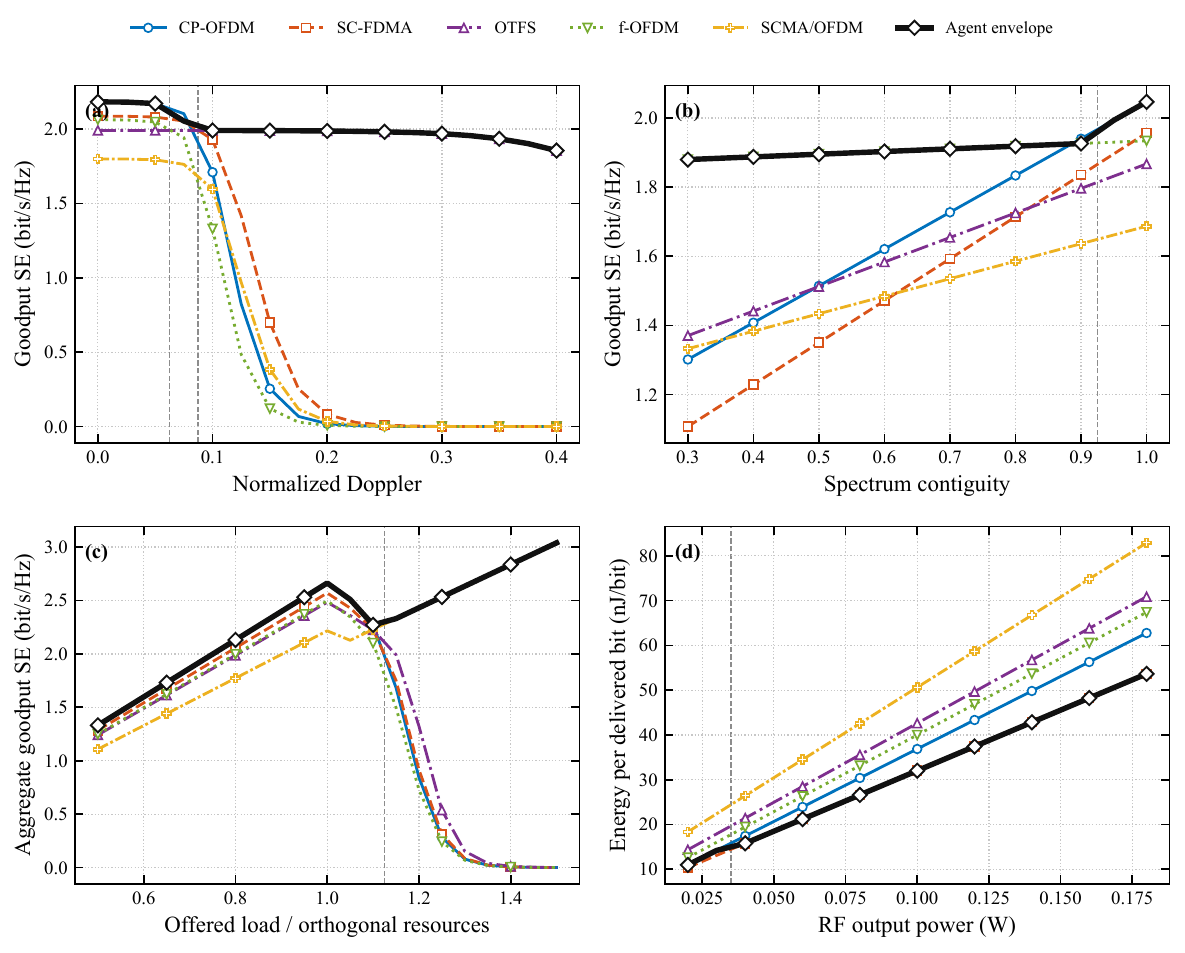}
    \caption{Level-3 continuous sweeps. The agent envelope changes tool as normalized Doppler, spectrum contiguity, offered load, or RF output power changes. Vertical dashed lines mark empirical switching points under the current models and gates.}
    \label{fig:l3}
\end{figure*}

The five formal operating points in Table~\ref{tab:l3results} select all five registered tools, but this one-to-one outcome is not preprogrammed. It follows from the current gates and weights. The continuous sweeps in Fig.~\ref{fig:l3} make that distinction testable. CP-OFDM is selected for normalized Doppler 0--0.05, SC-FDMA near 0.08, and OTFS from approximately 0.10 to 0.40. Filtered OFDM is selected for spectrum contiguity 0.30--0.90, returning to CP-OFDM near a fully contiguous band. CP-OFDM remains feasible up to normalized load 1.10; SCMA-over-OFDM is selected from 1.15 to 1.50. Under the present PA model, CP-OFDM is selected at 0.02--0.03 W RF output and SC-FDMA at 0.04--0.18 W.

These boundaries are empirical properties of the current channel distributions, power model, task weights, and hard gates. They are not standards constants. Altering the PA model, BLER abstraction, bandwidth, candidate implementation, or service weights should move the switching points. This sensitivity is a desired property: the purpose of the agent is to re-evaluate communication form when the relevant context changes.

\section{Discussion and Deployment Boundary}
\label{sec:discussion}

\subsection{Two Paradigm Transitions}

The results support the two transitions that motivate the paper. The first is a transition in \emph{how algorithms are produced}: manual model-based derivation is increasingly supplemented by foundation models that connect knowledge, code, simulation, and evaluators. The second is a transition in \emph{when algorithms can change}: an offline-selected implementation becomes a versioned runtime object that an agent may parameterize, replace, or reconfigure after executable verification. The second transition is the focus of the present experiments.

The three levels quantify the scale of that transition. Level~1 changes values inside a fixed graph. Level~2 changes the numerical algorithm used to realize a fixed receiver objective. Level~3 changes the transmitted representation or multiple-access chain. The progression is not a claim that a higher level is always better. When direct MMSE or CP-OFDM is the safest feasible action, the agent should preserve it. Intelligence includes the ability not to change.

\subsection{Safety and Implementation}

Four mechanisms separate controlled adaptation from arbitrary online mutation. First, every action belongs to a registered and typed library. Second, hard gates precede task utility. Third, both endpoints must agree on a Level-3 artifact and activation epoch. Fourth, monitoring, hysteresis, and fallback remain active after deployment. A future implementation can lower the versioned artifact to C/CUDA, FPGA overlays, or reconfigurable accelerator instructions, but the lowering and hardware mapping must themselves be verified.

The natural implementation is split between a non-real-time intelligence plane and a deterministic real-time control plane. The intelligence plane may use an LLM, search, simulation, and historical traces to enlarge or reprioritize the library. The real-time controller receives only compact manifests, calibrated metric models, thresholds, and policy tables. It executes bounded state estimation, gate evaluation, scheduling, and watchdog logic; no token generation is required at the symbol, sample, or channel-block rate. This separation makes worst-case timing analysis possible and allows the communications link to continue with the last-known-good artifact if the agent service, network connection, or accelerator manager is unavailable.

Hardware admission should be evidence-based at the artifact level. Software targets require binary hashes, ABI compatibility, memory isolation, and measured worst-case execution time. FPGA targets require signed bitstreams or overlays, static-shell compatibility, timing closure, resource bounds, and fixed-point regression. ASIC targets require an existing engine or microcode path and characterized voltage/frequency/thermal operating points. RF profiles require calibrated tuning ranges, EVM, ACLR, spurious-emission, and settling-time evidence. The deployment guard checks these properties before performance utility; a candidate with higher simulated rate is inadmissible if its implementation can miss a deadline, overflow local memory, violate a spectral mask, or leave no verified rollback path.

Monitoring must also be hierarchical. Fast hardware watchdogs observe overflow, residual growth, missed deadlines, thermal alarms, PLL unlock, and buffer under/overflow. Link-level monitors aggregate BLER, retransmission, EVM, CSI age, and queueing. The supervisory agent observes longer-term utility and distribution shift. A fast violation triggers local fallback without waiting for the LLM or global optimizer; repeated fallback events invalidate the manifest for that context and return evidence to the offline generation loop.

\subsection{Evidence Limitations}

The evidence has three principal limits. First, Level~1 uses coded baseband Monte Carlo and engineering models; it is not a standards-conformant BLER/HARQ experiment. Second, Level~2 uses correlated Rayleigh block fading, linear detection, hard QAM decisions, and analytical energy/latency proxies. It does not yet include 3GPP CDL channels, soft-output decoding, fixed-point overflow, memory contention, or measured ASIC/FPGA power. Third, Level~3 generates baseband samples for PAPR and leakage, but its BLER, latency, complexity, PA, and power models remain analytical. The experiments validate the adaptation mechanism and expose working regions; they do not certify a radio implementation.

The next steps are therefore concrete: add LDPC/Polar coding and soft outputs; replace analytical power by synthesis- or measurement-backed models; calibrate synchronization, EVM, and spectral masks; generate dual-endpoint implementations; and validate switching, negotiation, and rollback in an SDR hardware-in-the-loop platform. Foundation-model proposers can then expand the candidate library, while the same executable gates retain deployment authority.

\section{Conclusion}
\label{sec:conclusion}

This paper presented the \fabric{} as a mechanism for moving from fixed-algorithm execution to verifiable runtime communication adaptation. The central change is not a larger offline search alone. It is the combination of design intelligence with runtime intelligence: AI can assist the production of communication algorithms, and an agent can subsequently decide whether a parameter, receiver algorithm, or waveform remains appropriate under the current task, channel, spectrum, and hardware constraints. Level-1 experiments demonstrated low-risk parameter adaptation, Level-2 experiments established condition-dependent switching among Jacobi, PCG, and direct MMSE, and Level-3 experiments formed waveform and multiple-access switching regions across five registered tools. The results do not imply that one new waveform or solver should replace existing standards. They show that future communication systems can treat algorithms as explicit, versioned, and verifiable runtime objects whose modification scale is matched by an appropriate validation and deployment boundary.

\appendices

\section{Execution Budgets and Reproducibility}
\label{app:budget}

\footnotesize
Level~1 uses seed 20260811, four repetitions per candidate, 500 coded blocks, and 1200 PAPR blocks per repetition. Level~2 uses seed 20260818, 700 common-random-number channel blocks per formal scenario; its SNR sweep uses 500 blocks per point, $32\times8$ MIMO, 16-QAM, and transmit correlation 0.05. Zero empirical BER remains zero in JSON, while plots use $0.5/N_{\mathrm{bits}}$ only for logarithmic placement. Level~3 uses seed 20260818 for five formal scenes and four continuous sweeps. The frozen artifacts are \path{three_layer_results.json}, \path{layer2_receiver_adaptation.json}, \path{layer2_snr_sweep.json}, \path{layer3_waveform_adaptation.json}, and \path{layer3_waveform_sweeps.json} under \path{results/}; executable entry points with matching names are under \path{sim/} in the companion \path{agent-communication-demo} workspace.
\normalsize

\section{Complete Level-1 Candidate Evidence}

Table~\ref{tab:l1all} reports every generated Level-1 candidate. The selected row is the highest-utility candidate among those that pass the reliability and complexity gates; a lower-cost or higher-rate row is not eligible when its reliability gate fails.

\begin{table*}[t]
\caption{All Level-1 candidates. Rate is in Mb/s, latency in ms, and energy in nJ/bit. R/C denotes reliability/complexity gate.}
\label{tab:l1all}
\centering
\scriptsize
\begin{tabular}{llllllllrrrl}
\toprule
\textbf{Agent} & $M$ & $R_c$ & $\rho_p$ & \textbf{EQ} & $b_q$ & \textbf{BER} & \textbf{R/C} & \textbf{Rate} & \textbf{Latency} & \textbf{Energy} & \textbf{Selected} \\
\midrule
MCS & 4 & 1/2 & 0.08 & robust-MMSE & 8 & $8.152\!\times\!10^{-5}$ & Y/Y & 14.05 & 2387.93 & 121.93 & -- \\
MCS & 4 & 3/4 & 0.08 & robust-MMSE & 8 & $9.143\!\times\!10^{-4}$ & Y/Y & 21.08 & 1591.99 & 80.46 & -- \\
MCS & 16 & 1/2 & 0.08 & robust-MMSE & 8 & $2.739\!\times\!10^{-3}$ & Y/Y & 28.10 & 1194.02 & 60.27 & -- \\
MCS & 16 & 3/4 & 0.08 & robust-MMSE & 8 & $7.750\!\times\!10^{-3}$ & Y/Y & 42.16 & 796.05 & 39.70 & Y \\
MCS & 64 & 1/2 & 0.08 & robust-MMSE & 8 & $1.210\!\times\!10^{-2}$ & N/Y & 42.16 & 796.05 & 39.53 & -- \\
MCS & 64 & 3/4 & 0.08 & robust-MMSE & 8 & $3.606\!\times\!10^{-2}$ & N/Y & 63.24 & 530.73 & 26.71 & -- \\
\midrule
Tracking & 4 & 3/4 & 0.08 & MMSE & 8 & $7.363\!\times\!10^{-2}$ & N/Y & 4.14 & 190.11 & 162.02 & -- \\
Tracking & 4 & 3/4 & 0.08 & robust-MMSE & 8 & $6.909\!\times\!10^{-2}$ & N/Y & 4.14 & 190.12 & 166.84 & -- \\
Tracking & 4 & 3/4 & 0.16 & MMSE & 8 & $4.668\!\times\!10^{-2}$ & Y/Y & 3.78 & 208.20 & 178.56 & Y \\
Tracking & 4 & 3/4 & 0.16 & robust-MMSE & 8 & $5.046\!\times\!10^{-2}$ & Y/Y & 3.78 & 208.21 & 182.00 & -- \\
\midrule
Precision & 16 & 3/4 & 0.08 & robust-MMSE & 8 & $1.873\!\times\!10^{-3}$ & Y/Y & 18.40 & 57.14 & 13.44 & Y \\
Precision & 16 & 3/4 & 0.08 & robust-MMSE & 12 & $1.933\!\times\!10^{-3}$ & Y/Y & 18.40 & 57.14 & 14.61 & -- \\
\bottomrule
\end{tabular}
\end{table*}

\section{Complete Level-2 Solver Evidence}

Table~\ref{tab:l2all} retains the 24 common-random-number evaluations used by the slow-timescale policy. The direct solution is not automatically selected whenever it is feasible: the policy first removes infeasible candidates, then applies scenario weights to the remaining quality--cost tradeoff. The fast guard described in Algorithm~\ref{alg:l2} operates after this default selection.

\begin{table*}[t]
\caption{All Level-2 solver candidates. $r_{95}$ is P95 normalized residual; Div. is divergence rate; energy is nJ/vector and latency is $\mu$s/vector. F/M/H denote favorable, moderate, and harsh scenarios.}
\label{tab:l2all}
\centering
\scriptsize
\resizebox{\textwidth}{!}{%
\begin{tabular}{llllllllrrrl}
\toprule
\textbf{Scene} & \textbf{ID} & \textbf{Solver} & $T$ & $\omega$ & \textbf{BER} & \textbf{EVM} & $r_{95}$ & \textbf{Div.} & \textbf{Energy} & \textbf{Latency} & \textbf{Feasible/selected} \\
\midrule
F & J2 & Jacobi & 2 & .90 & 0 & 18.63\% & .286 & .09\% & 1.45 & .016 & N/-- \\
F & J3 & Jacobi & 3 & .90 & 0 & 10.76\% & .178 & 0 & 1.95 & .024 & Y/Y \\
F & J4 & Jacobi & 4 & .90 & 0 & 6.91\% & .114 & 0 & 2.45 & .032 & Y/-- \\
F & J6 & Jacobi & 6 & .82 & 0 & 4.33\% & .022 & 0 & 3.46 & .048 & Y/-- \\
F & P2 & PCG & 2 & -- & 0 & 13.40\% & .133 & 0 & 2.65 & .028 & N/-- \\
F & P3 & PCG & 3 & -- & 0 & 5.55\% & .047 & 0 & 3.58 & .042 & Y/-- \\
F & P5 & PCG & 5 & -- & 0 & 3.60\% & .004 & 0 & 5.44 & .070 & Y/-- \\
F & D & Direct & -- & -- & 0 & 3.58\% & 0 & 0 & 5.94 & .090 & Y/-- \\
\midrule
M & J2 & Jacobi & 2 & .90 & .3605 & 99.35\% & 3.188 & 96.29\% & 1.45 & .016 & N/-- \\
M & J3 & Jacobi & 3 & .90 & .2743 & 157.81\% & 6.126 & 96.40\% & 1.95 & .024 & N/-- \\
M & J4 & Jacobi & 4 & .90 & .4212 & 279.14\% & 11.903 & 96.69\% & 2.45 & .032 & N/-- \\
M & J6 & Jacobi & 6 & .82 & .3630 & 448.44\% & 19.114 & 85.96\% & 3.46 & .048 & N/-- \\
M & P2 & PCG & 2 & -- & .1731 & 47.40\% & .256 & .04\% & 2.65 & .028 & N/-- \\
M & P3 & PCG & 3 & -- & .09257 & 34.82\% & .135 & 0 & 3.58 & .042 & N/-- \\
M & P5 & PCG & 5 & -- & .02364 & 22.56\% & .038 & 0 & 5.44 & .070 & Y/Y \\
M & D & Direct & -- & -- & .01332 & 20.13\% & 0 & 0 & 6.04 & .090 & Y/-- \\
\midrule
H & J2 & Jacobi & 2 & .90 & .5945 & 466.36\% & 19.757 & 100\% & 1.98 & .020 & N/-- \\
H & J3 & Jacobi & 3 & .90 & .3570 & 1848.26\% & 89.425 & 100\% & 2.75 & .030 & N/-- \\
H & J4 & Jacobi & 4 & .90 & .6219 & 7650.32\% & 408.334 & 100\% & 3.51 & .040 & N/-- \\
H & J6 & Jacobi & 6 & .82 & .6235 & 71294.01\% & 4302.772 & 100\% & 5.04 & .060 & N/-- \\
H & P2 & PCG & 2 & -- & .1951 & 75.04\% & .272 & .24\% & 3.52 & .032 & N/-- \\
H & P3 & PCG & 3 & -- & .1339 & 66.35\% & .119 & 0 & 4.88 & .048 & N/-- \\
H & P5 & PCG & 5 & -- & .09395 & 60.08\% & .022 & 0 & 7.60 & .080 & Y/-- \\
H & D & Direct & -- & -- & .09043 & 59.53\% & 0 & 0 & 9.10 & .119 & Y/Y \\
\bottomrule
\end{tabular}}
\end{table*}

\section{Complete Level-3 Waveform Evidence}

Table~\ref{tab:l3all} evaluates every executable waveform tool in every formal scene. A row marked infeasible violates at least one reliability, ACLR, latency, power, or complexity gate. Large energy and latency values in failed high-mobility or overloaded orthogonal links are consequences of the current retransmission abstraction and should be read as failure indicators, not as hardware predictions.

\begin{table*}[t]
\caption{All Level-3 waveform/multiple-access candidates. SE is goodput bit/s/Hz, power is W, energy is nJ/bit, latency is ms, and $C$ is normalized processing complexity. B/U/M/F/A denote broadband downlink, battery uplink, high mobility, fragmented spectrum, and massive access.}
\label{tab:l3all}
\centering
\scriptsize
\resizebox{\textwidth}{!}{%
\begin{tabular}{llllllllrrrl}
\toprule
\textbf{Scene} & \textbf{Tool} & \textbf{BLER} & \textbf{SE} & \textbf{PAPR} & \textbf{ACLR} & \textbf{Power} & \textbf{Energy} & \textbf{Latency} & $C$ & \textbf{Feasible} & \textbf{Selected} \\
\midrule
B & CP-OFDM & $2.34\!\times\!10^{-4}$ & 2.320 & 10.34 & 18.49 & 1.427 & 30.75 & 723.17 & 1.00 & Y & Y \\
B & SC-FDMA & $2.15\!\times\!10^{-4}$ & 2.218 & 8.57 & 18.55 & 1.240 & 27.96 & 756.41 & 1.22 & Y & -- \\
B & OTFS & $1.58\!\times\!10^{-4}$ & 2.116 & 10.32 & 18.58 & 1.611 & 38.08 & 792.82 & 2.75 & N & -- \\
B & f-OFDM & $2.48\!\times\!10^{-4}$ & 2.192 & 10.33 & 22.66 & 1.506 & 34.33 & 765.23 & 1.75 & Y & -- \\
B & SCMA/OFDM & $1.63\!\times\!10^{-4}$ & 1.912 & 10.56 & 19.12 & 1.798 & 47.00 & 877.39 & 4.20 & N & -- \\
\midrule
U & CP-OFDM & $4.91\!\times\!10^{-3}$ & 1.902 & 10.34 & 18.49 & .547 & 28.76 & 220.57 & 1.00 & Y & -- \\
U & SC-FDMA & $3.78\!\times\!10^{-3}$ & 1.820 & 8.57 & 18.55 & .459 & 25.24 & 230.45 & 1.22 & Y & Y \\
U & OTFS & $1.55\!\times\!10^{-3}$ & 1.740 & 10.32 & 18.58 & .588 & 33.78 & 241.02 & 2.75 & N & -- \\
U & f-OFDM & $5.79\!\times\!10^{-3}$ & 1.796 & 10.33 & 22.66 & .564 & 31.42 & 233.60 & 1.75 & N & -- \\
U & SCMA/OFDM & $3.22\!\times\!10^{-3}$ & 1.570 & 10.56 & 19.12 & .636 & 40.53 & 267.18 & 4.20 & N & -- \\
\midrule
M & CP-OFDM & 1.00 & 0 & 10.34 & 18.49 & .858 & $2.03\!\times\!10^6$ & $4.97\!\times\!10^6$ & 1.00 & N & -- \\
M & SC-FDMA & 1.00 & 0 & 8.57 & 18.55 & .728 & $5.72\!\times\!10^5$ & $1.65\!\times\!10^6$ & 1.22 & N & -- \\
M & OTFS & $7.92\!\times\!10^{-3}$ & 1.976 & 10.32 & 18.58 & .935 & 47.32 & 106.13 & 2.75 & Y & Y \\
M & f-OFDM & 1.00 & 0 & 10.33 & 22.66 & .891 & $1.68\!\times\!10^7$ & $3.96\!\times\!10^7$ & 1.75 & N & -- \\
M & SCMA/OFDM & 1.00 & 0 & 10.56 & 19.12 & 1.021 & $1.47\!\times\!10^6$ & $3.01\!\times\!10^6$ & 4.20 & N & -- \\
\midrule
F & CP-OFDM & $5.38\!\times\!10^{-4}$ & 1.429 & 10.34 & 18.49 & .585 & 40.95 & 73.37 & 1.00 & N & -- \\
F & SC-FDMA & $4.37\!\times\!10^{-4}$ & 1.253 & 8.57 & 18.55 & .499 & 39.85 & 83.69 & 1.22 & N & -- \\
F & OTFS & $2.20\!\times\!10^{-4}$ & 1.456 & 10.32 & 18.58 & .643 & 44.20 & 72.05 & 2.75 & N & -- \\
F & f-OFDM & $6.16\!\times\!10^{-4}$ & 1.889 & 10.33 & 22.66 & .610 & 32.29 & 55.52 & 1.75 & Y & Y \\
F & SCMA/OFDM & $3.59\!\times\!10^{-4}$ & 1.393 & 10.56 & 19.12 & .706 & 50.68 & 75.27 & 4.20 & N & -- \\
\midrule
A & CP-OFDM & .989 & .010 & 10.34 & 18.49 & .491 & 5088.80 & 2718.54 & 1.00 & N & -- \\
A & SC-FDMA & .989 & .010 & 8.57 & 18.55 & .421 & 4244.24 & 2642.22 & 1.22 & N & -- \\
A & OTFS & .981 & .018 & 10.32 & 18.58 & .544 & 3108.27 & 1497.84 & 2.75 & N & -- \\
A & f-OFDM & .990 & .009 & 10.33 & 22.66 & .513 & 5469.40 & 2792.86 & 1.75 & N & -- \\
A & SCMA/OFDM & $1.06\!\times\!10^{-2}$ & 2.645 & 10.56 & 19.12 & .600 & 22.69 & 9.92 & 4.20 & Y & Y \\
\bottomrule
\end{tabular}}
\end{table*}

\bibliographystyle{IEEEtran}
\bibliography{references}

\end{document}